\documentclass[%
 preprint, 
 amsmath,amssymb,
 aps, physrev,
 showkeys,
]{revtex4-2}

\usepackage{graphicx}
\usepackage{dcolumn}
\usepackage{bm}

\usepackage{amsmath}

\usepackage{mathrsfs}
\usepackage{float}
\usepackage{multirow}
\usepackage{subfigure}
\usepackage{tabularx}
\usepackage{booktabs}
\usepackage{color}
\usepackage{enumitem}

\begin{document}

\preprint{APS/123-QED}

\title{\textbf{Intrinsic Incompatibility: Why Static Droplets Cannot Exist in Cahn-Hilliard-Navier-Stokes Systems} 
}%

\author{Jun Lai}
\email{Contact author: junlai@pku.edu.cn}
 \affiliation{%
 	School of Ocean Engineering, Guangzhou Maritime University, Guangdong 510725, P.R. China
 }%
%
%
%

\date{\today}

\begin{abstract}
Static droplets serve as fundamental benchmarks for interface-resolved simulations of two-phase flows. However, their accurate representation in phase-field models remains elusive due to persistent numerical artifacts. 
This work rigorously proves that
static droplets cannot exist in phase-field models governed by the Cahn-Hilliard-Navier-Stokes equations. Through equilibrium analysis of the governing equations, we demonstrate that
equilibrium necessitates uniform chemical potential ($\mu_{\phi}= 0$),
which nullifies the interfacial force ($\boldsymbol{F}=\boldsymbol{0}$),
enforcing a uniform pressure field ($\nabla p=\boldsymbol{0}$).
This directly contradicts the pressure jump required by Laplace’s law for a curved interface, proving mechanical equilibrium is impossible.
The results reveal an intrinsic incompatibility between non-flat equilibrium interfaces and the Cahn-Hilliard-Navier-Stokes system, 
provides a fundamental theoretical explanation for long-standing paradoxes such as
droplet shrinkage and parasitic currents.
This fundamental limitation applies universally to droplets/bubbles and necessitates re-evaluation of phase-field models for multiphase systems.
\end{abstract}

\maketitle


\section{\label{sec:Intro}Introduction}

Droplets serve as fundamental building blocks in multiphase flow physics and critical benchmarks for interface-resolved direct numerical simulations. Their geometric simplicity enables rigorous validation of numerical models, while their ubiquity in natural and industrial processes – from cloud microphysics to microfluidics – underscores practical significance~\cite{ovadnevaite2017surfaceNature,lohse2020NatureReviewsPhysics,GiuseppeNegro2023sciadv.adf8106,singh2024anomalousNatureCommunications,Chen2024PhysRevLett.133.028402,damak2022dynamicsScienceAdvances,allwayin2024locallyScience}. Consequently, accurately capturing static droplet behavior is a necessary condition for conducting reliable simulations of complex two-phase fluid flow.

The phase-field model, a prevalent diffuse-interface approach~\cite{Anderson1998DIFFUSEaa,lohse2020NatureReviewsPhysics,Demont_Stoter_van_2023JFM,ten_Eikelder_2024JFM,Mostafavi2025} for interface-resolved direct numerical simulations, describes interfacial dynamics through the convection-diffusion of an order parameter governed by the Cahn-Hilliard  equation~\cite{cahn1958free,cahn1959free,Espath_Sarmiento_Vignal_Varga_Cortes_Dalcin_Calo_2016JFM,LiShen2020sciadv.abb0597,Giovangigli2021PhysRevE.104.054109,Demont_Stoter_van_2023JFM,Frohoff2023PhysRevLett.131.107201,GiuseppeNegro2023sciadv.adf8106,Chen2024PhysRevLett.133.028402}.
Its ability to handle topological changes without explicit interface tracking has fueled widespread adoption in simulations~\cite{Padhan_Pandit_2025JFM} ranging from simple flow configurations~\cite{zu2013phase,zhang2018discrete,Busuioc_Kusumaatmaja_Ambruş_2020JFM,Demont_Stoter_van_2023JFM,Bachini_Krause_Nitschke_Voigt_2023,Kou_Salama_Wang_2023} to complex turbulent flow systems~\cite{scarbolo2013turbulence,komrakova2015numerical,roccon2017viscosity,Alberto2022sciadv.abp9561,lai2022systematic}. However, persistent numerical artifacts plague static droplet simulations, notably:
\begin{enumerate}[label=(\arabic*)]
	\item Droplet shrinkage~\cite{yue2007spontaneous,PhysRevE.100.061302}, historically attributed to finite interfacial thickness;
\item Parasitic currents (spurious velocities)~\cite{zu2013phase,li2025well}, often blamed on discrete force implementations;
\item Mass/volume non-conservation~\cite{WANG2015336,2016ALiYibao}, linked to numerical diffusion.
\end{enumerate}

Despite these empirical observations, a fundamental question remains unaddressed:
Is the phase-field model inherently compatible with the existence of static droplets?
Prior theoretical work has focused primarily on dynamic droplet behaviors. By assuming that the interfacial thickness is small compared to the droplet radius, and order parameter shifts uniformly to the bulk region, Yue {\it et al.}~(2007)~\cite{yue2007spontaneous} figured out the spontaneous shrinkage of droplets and critical drop radius that a droplet will eventually vanish, based on the mass conservation and energy decaying law for Cahn-Hilliard  equation.
Later, Zhang and Guo~(2019)~\cite{PhysRevE.100.061302} extended the research on critical radius to the problem involving solid walls. 
While prior studies analyzed the dynamic evolution and eventual disappearance of droplets under the assumption that a (quasi-)static configuration might transiently exist~\cite{yue2007spontaneous,PhysRevE.100.061302}, no work has rigorously established the existence or non-existence of a true static equilibrium solution satisfying all governing equations simultaneously.
Crucially, no study has reconciled the Cahn-Hilliard equation’s requirement for homogeneous chemical potential with the mechanical equilibrium condition imposed by Laplace’s law.

This work reveals a profound theoretical incompatibility.
We demonstrate that static droplets cannot exist as equilibrium solutions in the Cahn-Hilliard-Navier-Stokes framework. Our analysis proceeds as follows.
First, the governing equations for both incompressible and quasi-incompressible phase-field models are summarized in Section~\ref{sec :gover}. 
Then, we analyze the chemical potential and pressure jump of a static droplet in Section~\ref{sec: Theoretical analysis}.
The contradiction between the theoretical results and Laplace's law proves that static droplets cannot exist in this model.
The summary and conclusions are given in Section~\ref{Conclusion}.
The primary contribution of this work is the first rigorous proof of the theoretical impossibility of static droplets in the Cahn-Hilliard-Navier-Stokes model.

	\section{Theoretical framework}\label{sec :gover}

\subsection{Phase-field modeling fundamentals}\label{subsec :PFM}

The phase-field model~\cite{Anderson1998DIFFUSEaa,zhang2018discrete,liang2014phase,church2019high,zhang2019interface,Alberto2022sciadv.abp9561} is a diffuse-interface approach that resolves interfacial dynamics through continuous variation of an order parameter $\phi$.
This parameter distinguishes fluid phases ({\it e.g.}, $\phi_A=1$ for phase A, $\phi_B=0$ for phase B) while enabling natural handling of topological changes. The model’s thermodynamics are governed by a free-energy functional ${\cal F}(\phi,\nabla \phi)$ as~\cite{swift1996lattice,liu2003phase,jacqmin1996energy,yue2004diffuse,Chen2024PhysRevLett.133.028402}
\begin{equation}
{\cal F}(\phi,\nabla \phi)=\int_{\Omega}\underbrace{\psi(\phi)}_{\text{bulk energy density}}+\underbrace{\frac{\kappa}{2}|\nabla \phi|^{2}}_{\text{interfacial energy density}} d \Omega,\label{Fphi}
\end{equation}
where $\Omega$ is the volume of the considered system. 
$\kappa|\nabla \phi|^{2}/2$ penalizes interfacial gradients, driving mixing.
$\psi(\phi)$ is a double-well potential promoting phase separation,
\begin{equation}
\psi(\phi)=\beta\left(\phi-\phi_{A}\right)^{2}\left(\phi-\phi_{B}\right)^{2}.
\end{equation}
$ \kappa $ and $\beta$ are positive constants which uniquely determine the surface tension $\sigma$ and the interfacial thickness parameter $W$ together,
\begin{equation}
\sigma=\frac{\left|\phi_{A}-\phi_{B}\right|^{3}}{6} \sqrt{2 \kappa \beta},\quad
W=\frac{1}{\phi_{A}-\phi_{B}} \sqrt{\frac{8 \kappa}{\beta}}.
\end{equation}

The chemical potential $\mu_{\phi}$ is defined as the variation of the free-energy functional with respect to the order parameter,
\begin{equation}
{\mu_{\phi}=\frac{\delta {\cal F}}{\delta \phi}=\underbrace{4 \beta\left(\phi-\phi_{A}\right)\left(\phi-\phi_{B}\right)
	\left( \phi - { {\phi_{A} + \phi_{B} } \over 2} \right)}_{\text{bulk contribution}} \underbrace{-\kappa \nabla^{2} \phi}_{\text{interface contribution}}}.
\end{equation}
For a flat interface at steady state, 
\begin{equation}\label{muphi}
\mu_{\phi}=0,
\end{equation}
yielding the equilibrium profile
\begin{equation}\label{phi}
\phi(l)=\frac{\phi_{A}+\phi_{B}}{2}+\frac{\phi_{A}-\phi_{B}}{2} \tanh \left(\frac{2 l}{W}\right),
\end{equation}
where $l$ is the signed distance normal to the interface.

\subsection{Cahn-Hilliard-Navier-Stokes systems}\label{subsec :CHNS}

The evolution of the order parameter $\phi$ is governed by the Cahn-Hilliard  equation~\cite{cahn1958free,cahn1959free,DING20072078,WANG201541,WANG2015404,H2016Comparative,Busuioc_Kusumaatmaja_Ambruş_2020JFM,GiuseppeNegro2023sciadv.adf8106}, coupled with a convection term representing fluid advection. Two prevalent formulations exist for this convection term,
$\nabla \cdot(\phi \boldsymbol{u})$~\cite{WANG2015404,Espath_Sarmiento_Vignal_Varga_Cortes_Dalcin_Calo_2016JFM,zhang2019interface,Busuioc_Kusumaatmaja_Ambruş_2020JFM,Demont_Stoter_van_2023JFM,ZhangDa2024POF,H2016Comparative} or $\boldsymbol{u}\cdot\nabla \phi$~\cite{DING20072078,ABELS2012THERMODYNAMICALLY,yue2004diffuse,PhysRevE.87.023304,2017Error,Alberto2022sciadv.abp9561,GiuseppeNegro2023sciadv.adf8106,Mostafavi2025,Xiao2025POF}, 
where $\boldsymbol{u}$ is the fluid velocity. 
The corresponding evolution equations is
\begin{subequations}\label{evolphi}
	\begin{equation}\label{evolphi1}
	\frac{\partial \phi}{\partial t}+\nabla \cdot(\phi \boldsymbol{u})=\nabla \cdot \left(  M \nabla \mu_{\phi}\right) 
	\end{equation}      
	or
	\begin{equation}\label{evolphi2}
	\frac{\partial \phi}{\partial t}+\boldsymbol{u}\cdot\nabla \phi=\nabla \cdot \left(  M \nabla \mu_{\phi}\right) ,
	\end{equation}      
\end{subequations}
where $t$ is the evolution time, 
$M$ is the mobility. 

The density is typically defined as a function of order parameter, {\it i.e.}, $\rho=\rho\left( \phi\right) $. This functional dependence introduces a fundamental incompatibility between the conventional divergence-free condition for incompressible flows and the continuity equation in the vicinity of the interface~\cite{LaiJunPOF2025}. The incompressible model enforces a solenoidal velocity field throughout the entire domain~\cite{PhysRevE.97.033309,Espath_Sarmiento_Vignal_Varga_Cortes_Dalcin_Calo_2016JFM,Alberto2022sciadv.abp9561,Mostafavi2025,Demont_Stoter_van_2023JFM,Xiao2025POF},
\begin{equation}\label{eqmass1}
\nabla \cdot \boldsymbol{u}=0.
\end{equation}
Consequently, the continuity equation is generally violated near the interface~\cite{LaiJunPOF2025},
\begin{equation}\label{eqmass2}
\frac{\partial \rho}{\partial t}+\nabla \cdot(\rho \boldsymbol{u})=\frac{d\rho}{d\phi} \nabla \cdot \left(  M \nabla \mu_{\phi}\right)
\neq 0 \quad\text{(near interface)}.
\end{equation}
The quasi-incompressible model~\cite{PhysRevE.93.043303,zhang2018discrete}
prioritizes strict mass conservation~\cite{2017Improved,Busuioc_Kusumaatmaja_Ambruş_2020JFM,Giovangigli2021PhysRevE.104.054109,ZhangDa2024POF,lai2022systematic},
\begin{equation}\label{eqmass3}
\frac{\partial \rho}{\partial t}+\nabla \cdot(\rho \boldsymbol{u})=0.
\end{equation}
Under this constraint, the velocity field is necessarily compressible near the interface~\cite{LaiJunPOF2025},
\begin{subequations}\label{eqmass4}
	\begin{equation}\label{eqmass4.1}
	\nabla \cdot \boldsymbol{u}=\frac{d\rho/d\phi}{\phi d\rho/d\phi-\rho}\nabla \cdot \left(  M \nabla \mu_{\phi}\right)\neq 0 \quad\text{(near interface)}
	\end{equation}
	or
	\begin{equation}\label{eqmass4.2}
	\quad \nabla \cdot \boldsymbol{u}=\frac{d\rho/d\phi}{-\rho}\nabla \cdot \left(  M \nabla \mu_{\phi}\right)\neq 0 \quad\text{(near interface)}
	\end{equation}
\end{subequations}
corresponding to Eqs.~(\ref{evolphi1}) and~(\ref{evolphi2}), respectively. 
The incompressible model cannot always satisfy the continuity equation near the interface, while the quasi-incompressible model is designed to satisfy the continuity equation in the whole domain.
Note that while the specific form of the convection term (Eqs.~(\ref{evolphi1}) or~(\ref{evolphi2})) influences the details of the quasi-incompressible constraint (Eqs.~\eqref{eqmass4.1} or~\eqref{eqmass4.2}), our subsequent equilibrium analysis (Section~\ref{sec: Theoretical analysis}) is independent of this choice, as will be shown.
A common choice for the density function is a linear interpolation~\cite{Xiao2025POF} between the pure phase densities $\rho_{A}$ and $\rho_{B}$, 
\begin{equation}
\rho=\frac{\phi-\phi_{B}}{\phi_{A}-\phi_{B}} \rho_{A}+\frac{\phi_{A}-\phi}{\phi_{A}-\phi_{B}} \rho_{B},
\end{equation}
The required derivatives are then
\begin{equation}
\frac{d\rho}{d\phi}=\frac{\rho_{A}-\rho_{B}}{\phi_{A}-\phi_{B}},\quad
\frac{d\rho/d\phi}{\phi d\rho/d\phi-\rho}=-\frac{\rho_{A}-\rho_{B}}{\phi_{A}\rho_{B}-\phi_{B}\rho_{A}},
\end{equation}
providing explicit expressions for the right-hand sides of Eqs.~\eqref{eqmass2},~\eqref{eqmass4.1} and~\eqref{eqmass4.2}.

The momentum equation is 
\begin{equation}
\frac{\partial(\rho \boldsymbol{u})}{\partial t}+\nabla \cdot(\rho \boldsymbol{u} \boldsymbol{u})=-\nabla p+\nabla \cdot\left[\mu\left(\nabla \boldsymbol{u}+ \boldsymbol{u}\nabla\right)\right]+ \boldsymbol{F},\label{EqMom}
\end{equation}
where $p$ is the pressure, $\mu=\mu\left( \phi\right) $ is the dynamic viscosity~\cite{LaiJunPOF2025}. The force $\boldsymbol{F}$ represents the interfacial force. Two common expressions for the force are used in the literature,  
\begin{equation}\label{eqF}
\boldsymbol{F}=-\phi\nabla\mu_{\phi}~\cite{2005A,WANG201541}\quad \text{or}\quad \boldsymbol{F}=\mu_{\phi}\nabla\phi~\cite{DING20072078,2017Improved,PhysRevE.97.033309,2005A}.
\end{equation}
Other body forces are not considered in the static droplet analysis.

Collectively, the Cahn-Hilliard equation (Eq.~\eqref{evolphi}), the mass conservation constraint (incompressible model or quasi-incompressible model), and the momentum equation (Eq.~\eqref{EqMom}) constitute the governing equations for two-phase flow within the Cahn-Hilliard-Navier-Stokes framework.

\section{Theoretical analysis of static droplets in Cahn-Hilliard-Navier-Stokes systems}\label{sec: Theoretical analysis}

\subsection{Chemical potential at equilibrium}\label{subsec :muphi}

For a static flow field at an equilibrium state, the transient term and the convection term vanish, reducing the Cahn-Hilliard equation (Eqs.~(\ref{evolphi1}) and~(\ref{evolphi2})) to
\begin{equation}\label{muphieq}
\nabla \cdot \left(  M \nabla \mu_{\phi}\right) =0.
\end{equation}
Therefore, the equations for the velocity field (Eqs.~\eqref{eqmass1},~\eqref{eqmass2},~\eqref{eqmass3}, and~\eqref{eqmass4}) are inherently satisfied and need no further consideration here.   

While a flat interface admits the solution $\mu_{\phi}=0$ (Eq.~\eqref{muphi}) trivially satisfying Eq.~\eqref{muphieq}, we now rigorously examine the case of a static droplet. The boundary condition at infinity, corresponding to a pure single-phase fluid, requires
\begin{equation}\label{muphieqbc}
\mu_{\phi} \left(  \infty\right) =0.
\end{equation} 
We claim that for any static droplet at equilibrium, the chemical potential must vanish identically throughout the domain, {\it i.e.},
Eq.~\eqref{muphi} is also satisfied, 
\begin{equation}\label{muphidrop}
\mu_{\phi}(\boldsymbol{x})=0, \quad \forall\boldsymbol{x} \in \Omega.
\end{equation}
Now we prove it.

A simple case is that $M$ is a constant, which is used in most of the literature to simulate the two-phase flows~\cite{jacqmin1999calculation,liu2003phase,PhysRevE.100.061302,lai2022systematic,Alberto2022sciadv.abp9561,GiuseppeNegro2023sciadv.adf8106,ZhangDa2024POF,LaiJunPOF2025}.
Eq.~\eqref{muphieq} simplifies to a harmonic equation
\begin{equation}
\nabla^2 \mu_{\phi}=0.\label{EqCH2}
\end{equation}
Solutions to Eq.~\eqref{EqCH2} are harmonic functions. Given the boundary condition Eq.~\eqref{muphieqbc} and the requirement for $\mu_{\phi}$ to be bounded and smooth within the finite domain containing the droplet, the Maximum Principle for harmonic functions dictates that $\mu_{\phi}$ must attain its maximum and minimum values on the boundary. Since Eq.~\eqref{muphieqbc} is the sole boundary value specification, the unique solution is Eq.~\eqref{muphidrop}.

In general, $M$ can be a function of $\phi$, then Eq.~\eqref{evolphi} is called singular Cahn-Hilliard equation~\cite{Shen_Yang_Wang_2013,BaoJin2025}.
Since the static droplet is spherically symmetric, Eq.~\eqref{muphieq} can be written as
	\begin{equation}\label{muphieqr}
	\frac{1}{r^{D-1}} \frac{d}{dr} \left( r^{D-1} M(r) \frac{d\mu_{\phi}(r)}{dr} \right) = 0 ,
	\end{equation}
	where $D$ represents the dimension of the space. $D=2,3$ for 2D and 3D droplets, respectively, with $r$ denotes the radial direction in polar coordinates and spherical coordinates.
	Integrating Eq.~\eqref{muphieqr} twice yields the general solution,
	\begin{equation}\label{muphiC}
	\mu_{\phi}(r) = C_1 \int \frac{dr}{r^{D-1} M(r)} + C_2,
	\end{equation}
	where $C_1$ and $C_2$ are two integration constants.
	Within the droplet interior, the order parameter $\phi$ (and consequently $M(\phi)$) is approximately constant. The integral term in Eq.~\eqref{muphiC} exhibits a singularity at the droplet center. To ensure $\mu_{\phi}$ remains bounded and physically meaningful at the center, the constant $C_1$ must be zero.
	Then $C_2=0$ due to the boundary condition Eq.~\eqref{muphieqbc}.
	Thus, Eq.~\eqref{muphidrop} is satisfied.
	
	Therefore, for both constant and concentration-dependent mobility formulations, the chemical potential must vanish everywhere, $\mu_{\phi}\equiv 0$, for any static droplet at equilibrium.

\subsection{Pressure distribution and mechanical equilibrium contradiction}\label{subsec :p}

At equilibrium, the static flow field ($\partial/\partial t=0, \boldsymbol{u}=\boldsymbol{0}$) simplifies the momentum equation (Eq.~\eqref{EqMomeq}) to a balance between pressure gradients and interfacial force,
\begin{equation}\label{EqMomeq}
\nabla p=\boldsymbol{F},
\end{equation}
This equation implies that any pressure variations must be sustained by the interfacial force.

However, Subsection~\ref{subsec :muphi} established that $\mu_{\phi}\equiv 0$ everywhere for a static droplet (Eq.~\eqref{muphidrop}). Substituting this critical result into the common interfacial force expressions (Eq.~\eqref{eqF}) yields
\begin{equation}\label{eqF0}
\boldsymbol{F}\equiv\boldsymbol{0}.
\end{equation}
Consequently, the fundamental mechanism responsible for generating surface tension effects, the interfacial force $\boldsymbol{F}$, is absent at equilibrium.

Combining Eqs.~\eqref{EqMomeq} and~\eqref{eqF0} dictates that the pressure gradient must also vanish,
\begin{equation}\label{EqMomeq0}
\nabla p=\boldsymbol{0},
\end{equation}
This necessitates a uniform pressure distribution across the entire flow field, encompassing both the droplet interior and the surrounding fluid.

This uniform pressure state fundamentally contradicts the mechanical equilibrium condition for a curved fluid-fluid interface, as described by Laplace's law. Laplace's law requires a pressure jump $\delta p$ across the interface, proportional to the curvature ${\cal K}$ and the surface tension $\sigma$,
\begin{equation}
\delta p={\cal K}\sigma=\frac{\left( D-1\right) \sigma}{R},
\end{equation}
where $R$ is the radius of the static droplet.
A uniform pressure field is incompatible with this required pressure jump ($\delta p>0$) for any droplet with nonzero surface tension $\sigma$ and curvature ${\cal K}$.

This profound contradiction reveals a core theoretical limitation.
Within the Cahn-Hilliard-Navier-Stokes framework, mechanical equilibrium ($\nabla p =\boldsymbol{0}$)
enforced by the vanishing interfacial force ($\boldsymbol{F}=\boldsymbol{0}$ due to $\mu_{\phi}\equiv 0$) is irreconcilable with the non-uniform pressure distribution mandated by Laplace's law for a curved interface. Consequently,
no static droplet exhibiting a curved interface can exist as an equilibrium solution of the Cahn-Hilliard-Navier-Stokes equations.

This inherent incompatibility at equilibrium provides a unified theoretical foundation for at least three persistent numerical artifacts in phase-field simulations of static droplets.
(1) Droplet shrinkage and vanishing occur as the system attempts to minimize the free energy and the influence of chemical potential in the whole flow field;
(2) Parasitic currents manifest as unphysical circulatory flows generated by the system’s failed attempt to balance the nonexistent interfacial force with the pressure gradient required by Laplace’s law;
(3) Mass/volume non-conservation originates from the nullification of chemical potential gradients, which suppresses the stabilizing counter-fluxes needed to maintain droplet curvature against diffusive relaxation.
Collectively, these phenomena are not merely numerical artifacts but direct manifestations of the Cahn-Hilliard-Navier-Stokes system’s fundamental inability to satisfy both mechanical equilibrium and interfacial thermodynamics for curved interfaces.

\section{Summary and Conclusions}\label{Conclusion}
This study has established a fundamental theoretical limitation and pointed out a long-standing paradox of the phase-field model in describing static droplets within the Cahn-Hilliard-Navier-Stokes framework. Through rigorous analysis of the governing equations at equilibrium, we demonstrate that static droplets cannot exist as stable configurations
in systems governed by the Cahn-Hilliard equation. 

The contradiction arises from two core findings:
(1) For any static droplet at equilibrium, the chemical potential must vanish throughout the domain ($\mu_{\phi}\equiv 0$), regardless of mobility formulation (constant or $\phi$-dependent). 
(2) The condition ($\mu_{\phi}\equiv 0$) implies the absence of interfacial force. Consequently, the momentum equation reduces to $\nabla p=0$,
enforcing a uniform pressure field. This directly contradicts Laplace’s law ($\delta p \propto\sigma/R$), which requires a pressure jump sustained by surface tension.

Our theoretical analysis implies that the phase-field model with Cahn-Hilliard equation inherently prohibits non-flat equilibrium interfaces
in isolated droplet/bubble systems. 
This theoretical incompatibility explains persistent numerical artifacts ({\it e.g.}, parasitic currents) and non-physical droplet shrinkage observed in simulations.
The work resolves a long-standing paradox: While prior studies attributed droplet shrinkage to finite interfacial thickness or numerical errors, we prove it is
an intrinsic property of the Cahn-Hilliard-Navier-Stokes formulation at equilibrium.
It is noteworthy that although this paper exclusively addresses droplet dynamics, all conclusions are equally applicable to bubble systems, owing to their fundamental consistency in governing equations, interfacial dynamics, and numerical simulation approaches.

These conclusions apply universally to bubbles and droplets governed by the Cahn-Hilliard-Navier-Stokes equations, impacting the interpretation of benchmark simulations and multiphase flow physics. Future work should address this limitation through:
(1) Development of thermodynamically consistent model variants.
(2) Exploration of generalized chemical potential formulations.
(3) Re-assessment of numerical schemes for steady-state two-phase flows.
This work provides a foundational insight into the mathematical constraints of diffuse-interface models, urging caution in their application to static or quasi-static multiphase systems.


\begin{acknowledgments}
	This work has been supported by Science and Technology Projects in Guangzhou (SL2024A04J01991),
	Tertiary Education Scientific research project of Guangzhou Municipal Education Bureau (2024312534),
	and Start-up Fund for
	Talent Introduction at Guangzhou Jiaotong University (K42024035).
\end{acknowledgments}

\bibliography{apssamp}

\end{document}